# Four-layer nanocomposite structure as an effective optical waveguide switcher for near-IR regime

**I S Panyaev[1], N N Dadoenkova[1,2], Yu S Dadoenkova[1,2,3], I A Rozhleys[4], M Krawczyk[5], I L Lyubchanskii[2], and D G Sannikov[1]**

[1] Ulyanovsk State University, 432017 Ulyanovsk, Russian Federation
[2] Donetsk Physical and Technical Institute of the National Academy of Sciences of Ukraine, 83114 Donetsk, Ukraine
[3] Novgorod State University, 173003 Veliky Novgorod, Russian Federation
[4] National Research Nuclear University MEPhI, 115409 Moscow, Russian Federation
[5] Faculty of Physics, Adam Mickiewicz University in Poznań, 61–614 Poznań, Poland

E-mail: sannikov-dg@yandex.ru

Abstract. We present a theoretical study of the dispersion and energy properties of the eigenwaves (TE- and TM- modes) in a four-layer structure composed of a magneto-optical yttrium iron garnet guiding layer on a dielectric substrate covered by a planar nanocomposite guiding multilayer. The bigyrotropic properties of yttrium-iron garnet are taken into account for obtaining the dispersion equation and an original algorithm for the guided modes identification is proposed. We demonstrated the polarization switching of TE- and TM-modes dependent on the geometrical parameters of the guiding layers. The dispersion diagrams and field profiles are used to illustrate the change of propagation properties with variation of the multilayer thickness ratio of the nanocomposite's layers. The energy flux distributions across the structure are calculated and the conditions of the optimal guiding regime are obtained. The power switching ratio in the waveguide layers of about 6 dB for the wavelength range of 100 nm is shown to be achieved.

## 1. Introduction

Transparent in optical and infrared regimes multilayer structures have been the subjects of intensive theoretical and experimental research because of their wide applications in integrated optics, optoelectronics and photonics during more than forty years [1–11]. The three-layer waveguide usually provides the interconnections in an integrated optical circuits, and it is thus a multilayer structure which forms the actual components of an active or passive devices [12]. The history of investigations of the optical four-layer structures dates back to the late 60s - early 70's when the theory of four-layer isotropic waveguides had been developing [13–15], where a prism was used as a forth (cladding) medium, and an air gap between the prism and the film was considered as a coupling zone, where the fields are evanescent waves. The early experimental measurements of the waveguide modal characteristics have been carried out by Tien *et al*. [16] and Sun and Muller [17]. About the same time there was a rapid growth of the heterostructure laser technologies, particularly, the efficient large optical cavity injection lasers were fabricated [18]. A rich variety of the effects and applications of the four-layer structures were studied, for example, the TE/TM polarizers [19], leaky modes lasers [20], optical waveguide lenses and tapered couplers [21], *etc*. Presence of an extra layer in a conventional three-layer waveguide structure leads to an enhancement of light control options. The metal or semiconductor thin film can be used as the fourth layer, and this means that the absorption losses and the mode interaction should be considered [22,23]. The S-shaped four-layer waveguides were investigated in [24]. The theory had its further mathematical development in obtaining the general (normalized) dispersion relation and founding its application to the control of chromatic dispersion in a thin-film Luneburg lens [12,25]. The polarization effects as well as the combined effects of the waveguide coupling and losses were determined and illustrated in the devices using the silicon claddings [26]. The semiconductor-based four-layer structure has been studied and proposed as the basic building block for mid-infrared modulators and switches [27,28].

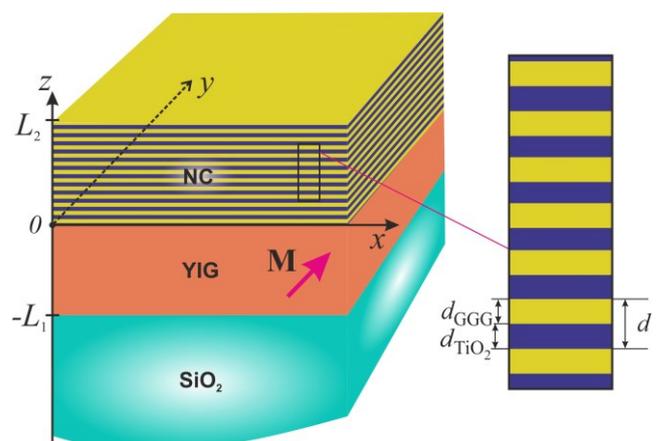

**Figure 1.** A schematic image of the four-layer structure: layer 1 – YIG; layer 2 – NC multilayer; layer 3 – SiO$_2$ substrate; layer 4 – air.

The investigation of gyromagnetic and gyroelectric properties of the multilayered nanocomposite (NC) structures is of great interest due to their specific properties which make them suitable for wide applications [29]. The optical isolator based on the leaky yttrium iron garnet (YIG) $Y_3Fe_5O_{12}$ waveguide was designed using the transfer matrix formulation [30].

The investigation of the multimode four-layer waveguide structures by modified m-line spectroscopy method was presented in [31]. During the last decade different multilayered waveguiding structures had been studied: the magnetic photonic crystals [32–34], nanocomposite multilayers and photonic crystal waveguides [35], silicon-based hybrid gap surface plasmon polariton waveguides [36], nanophotonic and plasmonic waveguides [37] and multilayer graphene waveguides [38]. Besides, four-layer structures due to the enhanced field confinement in guiding layer [39,40] can be effectively used in medical applications such as label free biosensors.

In this paper, we investigate theoretically the dispersion and polarization properties of propagating TE- and TM-waves in the hybrid four-layer magneto-optical (MO) nanocomposite waveguide. The multilayered structure is composed of a magneto-optical layer on dielectric substrate covered by multilayered dielectric NC medium. In Section 2, using the effective medium approach [41,42], we obtain the dispersion equation for the multilayered waveguiding structure taking into account bigyrotropic properties of the MO layer and features of the NC topology and propose an algorithm, which unlike the 4×4 matrix method [43] allows to identify the guided modes and to obtain their electric and magnetic fields distributions across the structure. In Section 3, we perform the numerical computations for different dispersion dependencies in the conventional telecommunication frequency regimes and present the cross-structure flux distributions. Some possible applications of the structure under consideration are discussed in Section 4. In Section 5 we summarize the paper results.

## 2. Theoretical analysis of the complex waveguide structure

*2.1. Geometry of the waveguide structure*

Let us consider the four-layered waveguide structure consisting of a magneto-optical YIG film of a thickness $L_1$ on the thick $SiO_2$ substrate, and covered by a one-dimensional NC multilayer of a thickness $L_2$, and air playing a role of a cladding. The NC multilayer with the period $d = d_{GGG} + d_{TiO_2}$ is formed by alternating nanolayers of gadolinium-gallium garnet (GGG) $Gd_3Ga_5O_{12}$ and titanium oxide $TiO_2$ with the corresponding thicknesses $d_{GGG}$ and $d_{TiO_2}$. We consider the geometry with the layers of the planar structure $(SiO_2/YIG/(GGG/TiO_2)^N/air$ (where $N$ is the number of the NC periods) located in the *xy*-plane, and the *z*-axis is perpendicular to the interfaces, as shown in figure 1. The electromagnetic wave propagates along the *x*-axis, and the magnetization vector **M** is transverse to the wave propagation direction and is directed along the *y*-axis. We assume that in all the layers of the waveguide system (YIG, GGG and $TiO_2$) the electromagnetic radiation propagates along the *x*-axis, exponentially decaying in depth of the substrate and air along the *z*-axis. The considered geometry corresponds to the transverse magneto-optical configuration, and in this case the electromagnetic waves in the YIG layer split into independent TE- and TM-modes [29].

*2.2. Material parameters of the constituting layers*

It is well known that the YIG is transparent in the near infra-red (IR) regime [29,44] and exhibits bigyrotropic properties, so both dielectric permittivity $\hat{\varepsilon}_{YIG}$ and magnetic permeability $\hat{\mu}_{YIG}$ tensors contain the nonzero off-diagonal components [45]. For the case of the magnetization **M** along the *y*-axis, in the linear MO approximation, the permittivity and permeability tensors have the following form [43]:

$$\hat{\varepsilon}_{YIG} = \begin{pmatrix} \varepsilon_1 & 0 & i\varepsilon_a \\ 0 & \varepsilon_1 & 0 \\ -i\varepsilon_a & 0 & \varepsilon_1 \end{pmatrix}, \quad \hat{\mu}_{YIG} = \begin{pmatrix} \mu_1 & 0 & i\mu_a \\ 0 & \mu_1 & 0 \\ -i\mu_a & 0 & \mu_1 \end{pmatrix}. \tag{1}$$

From the Maxwell's equations we obtain the following relationships between the electric and magnetic fields components of the electromagnetic wave:

$$\begin{aligned} H'_x + i\beta H_z &= ik_0\varepsilon_0 E_y, \\ E'_y &= ik_0(\mu_1 H_x + i\mu_a H_z), \\ \beta E_y &= k_0(\mu_1 H_z - i\mu_a H_x), \\ H'_y &= -ik_0(\varepsilon_1 E_x + i\varepsilon_a E_z), \\ \beta H_y &= k_0(i\varepsilon_a E_z - \varepsilon_1 E_z), \\ i\beta E_z + E'_x &= -ik_0\mu_0 H_y, \end{aligned} \tag{2}$$

where $k_0 = \omega/c$ is a wave vector of the electromagnetic wave in air ($\omega$ is the angular frequency, and $c$ is the speed of light in vacuum), and $\beta$ is a propagation constant (wave vector component along the x-axis). The permittivity tensors for nonmagnetic dielectrics GGG and TiO$_2$ are diagonal $\hat{\varepsilon}_{GGG} = \varepsilon_{GGG}\hat{I}$ and $\hat{\varepsilon}_{TiO_2} = \varepsilon_{TiO_2}\hat{I}$, where $\hat{I} = \delta_{ik}$, and $\delta_{ik}$ is the Kronecker's delta. In the case of a large number of nanolayers $N$ with the thicknesses $d_{GGG}$ and $d_{TiO_2}$, satisfying the condition, $d_{GGG}, d_{TiO_2} \ll \lambda$ ($\lambda$ is the wavelength of the electromagnetic wave), one can use the effective medium method (the long-wave approximation), which allows to treat the NC medium as a uniaxial crystal with an effective permittivity in the form [46]:

$$\hat{\varepsilon}_{NC} = \begin{pmatrix} \varepsilon_{xx} & 0 & 0 \\ 0 & \varepsilon_{yy} & 0 \\ 0 & 0 & \varepsilon_{zz} \end{pmatrix} \tag{3}$$

with the tensor components

$$\varepsilon_{xx} = \varepsilon_{yy} = \frac{\Theta\varepsilon_{GGG} + \varepsilon_{TiO_2}}{\Theta + 1}, \quad \varepsilon_{zz} = \frac{\varepsilon_{GGG}\varepsilon_{TiO_2}(\Theta + 1)}{\Theta\varepsilon_{TiO_2} + \varepsilon_{GGG}}, \tag{4}$$

where $\Theta = d_{GGG}/d_{TiO_2}$ is the ratio of the GGG and TiO$_2$ layers thicknesses.

The substrate medium 3 (SiO$_2$) and the cladding medium 4 (air) have the scalar permittivities $\varepsilon_3$, $\varepsilon_4$ and permeabilities $\mu_3 = \mu_4 = 1$. The magnetic permeabilities of GGG, TiO$_2$ and, therefore, the permeability of NC, are equal to unity ($\mu_2 = 1$).

*2.3. Profile functions and dispersion relation*
Solving the Maxwell's equations with the given material parameters of all the layers, one can obtain the profile functions of the electric and magnetic fields for TE-and TM-modes. The field of the electromagnetic wave, propagating along the x-axis, is

$$\mathbf{F}(x,z) = \mathbf{F}(z)e^{i\beta x}, \tag{5}$$

The tangential components of the vector profile function $\mathbf{F}(z)$ (the electric field component $E_y$ for TE-mode and the magnetic field component $H_y$ for TM-mode) have the form:

$$F_y(z) = A \cdot \begin{cases} \cos h_1 z + C_1 \sin h_1 z, & -L_1 \leq z \leq 0, \\ \cos h_2 z + C_2 \sin h_2 z, & 0 \leq z \leq L_2, \\ \left[\cos h_1 L_1 - C_1 \sin h_1 L_1\right]e^{p(z+L_1)}, & z \leq -L_1, \\ \left[\cos h_2 L_2 + C_2 \sin h_1 L_1\right]e^{-q(z-L_2)}, & z \geq L_2. \end{cases} \tag{6}$$

Here $A$ is the normalized amplitude which can be calculated by integrating the Poynting vector's longitudinal component [47], the coefficients $C_1$ and $C_2$ are determined as:

$$C_1 = \frac{\delta p - h_1 \tan h_1 L_1 - \beta v}{\delta p \tan h_1 L_1 + h_1 - \beta v \tan h_1 L_1}, \quad C_2 = \frac{h_2 \tan h_2 L_2 - \sigma q}{h_2 + \sigma q \tan h_2 L_2}, \quad (7)$$

and the transverse components of the wave vector in each layer are defined as:

$$\begin{cases} h_1^2 = k_0^2 \varepsilon_1 \mu_\perp - \beta^2 & \text{(TE-modes),} \\ h_1^2 = k_0^2 \mu_1 \varepsilon_\perp - \beta^2 & \text{(TM-modes),} \end{cases}$$

$$\begin{cases} h_2^2 = k_0^2 \varepsilon_{yy} \mu_2 - \beta^2 & \text{(TE-modes),} \\ h_2^2 = k_0^2 \varepsilon_{xx} \mu_2 - \frac{\varepsilon_{xx}}{\varepsilon_{zz}} \beta^2 & \text{(TM-modes),} \end{cases} \quad (8)$$

$$\begin{cases} p^2 = k_0^2 \varepsilon_3 - \beta^2, \\ q^2 = \beta^2 - k_0^2 \varepsilon_4, \end{cases}$$

where $\sigma = \mu_2/\mu_4$, $\tau = \mu_1/\mu_2$, $\delta = \mu_1/\mu_3$, $v = \mu_a/\mu_1$ for TE-mode, and $\sigma = \varepsilon_{zz}/\varepsilon_4$, $\tau = \varepsilon_1/\varepsilon_{zz}$, $\delta = \varepsilon_1/\varepsilon_3$, $v = \varepsilon_a/\varepsilon_1$ for TM-mode, with $\mu_\perp = \mu_1 - \mu_a^2/\mu_1$ and $\varepsilon_\perp = \varepsilon_1 - \varepsilon_a^2/\varepsilon_1$ being the transverse permeability and permittivity, respectively. The dispersion equation which connects the propagation constant $\beta$ of the corresponding waveguide mode with the radiation frequency and waveguide structure parameters can be obtained from the continuity conditions for the tangential field components at the boundaries of the media and can be written as

$$\left[\delta\tau\, p h_2^2 + \sigma q h_1^2 + \beta v\left(\beta v\sigma q - \delta\sigma p q - \tau h_2^2\right)\right]\tan h_1 L_1 \cdot \tan h_2 L_2 +$$
$$h_2\left[h_1^2 - \delta\sigma\tau pq + \beta v\left(\beta v + \tau\sigma q - \delta p\right)\right]\tan h_1 L_1 + \quad (9)$$
$$h_1\left(\tau h_2^2 - \delta\sigma pq\right)\tan h_2 L_2 - h_1 h_2\left(\delta p + \sigma\tau q\right) = 0.$$

Another form of the dispersion equation for a non-gyrotropic media is obtained in Refs. [16,17] and the similar one is given in Refs. [48,49]. In the absence of the NC layer $(L_2 = 0)$, the Eq. (9) transforms into a well-known dispersion equation for a three-layer waveguide structure (see, e.g. [1,50]).

*2.4. Guiding mode identification*

In order to determine the number of nodes of the electric and magnetic fields for TE- and TM-modes, we rewrite the profile functions (6) for the layers 1 and 2 in the following form:

$$\begin{cases} F_y(-L_1 \leq z \leq 0) = A \cdot \text{sign}(C_1)\sqrt{1+C_1^2}\sin(h_1 z + \varphi_1), \\ F_y(0 \leq z \leq L_2) = A \cdot \text{sign}(C_2)\sqrt{1+C_2^2}\sin(h_2 z + \varphi_2). \end{cases} \quad (10)$$

Here $\varphi_1 = \arctan(1/C_1)$ and $\varphi_2 = \arctan(1/C_2)$ are the initial phases. The number of nodes can be defined as follows. In the chosen layer, for example, the layer 1 (YIG), if the initial phase $\varphi_1$ satisfies the condition $0 \leq \varphi_1 \leq |h_1 L_1| - \pi \cdot \{|h_1 L_1|/\pi\}_{\text{int}}$ (the subscript "int" denotes the integer part), then the number of nodes $M_1$ in the layer is equal to $M_1 = \{|h_1 L_1|/\pi\}_{\text{int}} + 1$. In all other cases the number of nodes is less by 1: $M_1 = \{|h_1 L_1|/\pi\}_{\text{int}}$. The number of nodes in the layer 2 (NC), $M_2$ is defined in the similar way:

$$\begin{cases} M_2 = \{h_2 L_2/\pi\}_{\text{int}} + 1, & \text{if } \pi \cdot \{h_2 L_2/\pi\}_{\text{int}} - h_2 L_2 \leq \varphi_2 \leq 0, \\ M_2 = \{h_2 L_2/\pi\}_{\text{int}}, & \text{in all other cases.} \end{cases} \quad (11)$$

Then, the order of the mode is determined as:

$$M = M_1 + M_2. \quad (12)$$

**3. Numerical results and discussion**

For the numerical analysis of the results obtained above, we take into account the dispersion of refractive indexes (the Sellmeier's equations) of YIG, SiO$_2$, GGG:

$$n_j^2 = 1 + \frac{A_1 \lambda^2}{\lambda^2 - B_1^2} + \frac{A_2 \lambda^2}{\lambda^2 - B_2^2} + \frac{A_3 \lambda^2}{\lambda^2 - B_3^2}. \tag{13}$$

**Table 1.** Sellmeier coefficients in Eq.(13).

| Material | $A_1$ | $A_2$ | $A_3$ | $B_1$ (μm) | $B_2$ (μm) | $B_3$ (μm) |
|---|---|---|---|---|---|---|
| YIG [51] | 3.739 | 0.79 | - | 0.28 | 10 | - |
| GGG [52] | 1.7727 | 0.9767 | 4.9668 | 0.1567 | 0.01375 | 22.715 |
| SiO$_2$ [53] | 0.6961663 | 0.4079426 | 0.8974794 | 0.0684043 | 0.1162414 | 9.896161 |

As for TiO$_2$, its dispersion is given by formula $n_{\text{TiO}_2}^2 = 5.913 + \frac{0.2441}{\lambda^2 - 0.0803}$ [54].

For GGG, TiO$_2$ and SiO$_2$ layers we obtain the following wavelength-dependent dielectric permittivities: $\varepsilon_{\text{GGG}}(\lambda) = n_{\text{GGG}}^2(\lambda)$, $\varepsilon_{\text{TiO}_2}(\lambda) = n_{\text{TiO}_2}^2(\lambda)$, $\varepsilon_3(\lambda) = n_{\text{SiO}_2}^2(\lambda)$. For the YIG layer, the diagonal components of the dielectric permittivity tensor $\hat{\varepsilon}_{\text{YIG}}$ in (1) are assumed to be $\varepsilon_1 = n_{\text{YIG}}^2(\lambda)$. According to [55], the off-diagonal material tensor elements for YIG are $\varepsilon_a = -2.47 \cdot 10^{-4}$ and $\mu_a = 8.76 \cdot 10^{-5}$. The magnetic permeability components of YIG can be taken as $\mu_1 = 1$ for the considered frequency regime [45].

The refractive indices of the NC multilayer and YIG film can be obtained from the wave localization conditions $h_1 = 0$ and $h_2 = 0$ [see Eq. (8)]. Thus, for the NC multilayer $n_2^{\text{TE}} = \sqrt{\varepsilon_{yy}\mu_2}$, $n_2^{\text{TM}} = \sqrt{\varepsilon_{zz}\mu_2}$ and for the YIG film $n_1^{\text{TE}} = \sqrt{\varepsilon_\perp \mu_1}$ and $n_1^{\text{TM}} = \sqrt{\varepsilon_1 \mu_\perp}$. It should be noted, that in the NC multilayer the refractive indices for TE- and TM-modes differ considerably, and an inequality $n_2^{\text{TE}} > n_2^{\text{TM}}$ is satisfied throughout the entire wavelength range for all values of $\Theta$, as it follows from Eqs. (4). The difference between $n_2^{\text{TE}}$ and $n_2^{\text{TM}}$ varies with $\Theta$ and depends on the dielectric permittivities of the NC constituents. Also, the difference shows the anisotropy of optical properties of TE- and TM- modes which is only due to the nanostructuring of the waveguide layer 2. In the YIG film the refractive indices $n_1^{\text{TE}}$ and $n_1^{\text{TM}}$ are almost equal due to the small off-diagonal components of permittivity and permeability tensors in the considered wavelength range, so below we assume that $n_1^{\text{TE}} \approx n_1^{\text{TM}} \equiv n_1$.

In figure 2 we present the dispersion of the refractive indices calculated within $\lambda = (1 \div 6)$ μm for different values of the ratio between the thicknesses of GGG and TiO$_2$ layers constituting the NC multilayer: (a) $\Theta = 0.5$, (b) $\Theta = 0.62$, (c) $\Theta = 1$, and (d) $\Theta = 1.75$. The solid, dashed and dot-dashed lines denote $n_2^{\text{TE}}(\lambda)$, $n_2^{\text{TM}}(\lambda)$ and $n_1(\lambda)$, respectively.

As one can see from figure 2, for the fixed wavelength, one of the following conditions can be satisfied:

$$n_1 < n_2^{\text{TM}} < n_2^{\text{TE}}, \tag{14a}$$
$$n_2^{\text{TM}} < n_1 < n_2^{\text{TE}}, \tag{14b}$$
$$n_2^{\text{TM}} < n_2^{\text{TE}} < n_1. \tag{14c}$$

The dispersion curves of the waveguide modes propagating in both layers (YIG and NC) are located between the curves $n_1(\lambda)$ and $n_2^{\text{TM}}(\lambda)$ [or $n_1(\lambda)$ and $n_2^{\text{TE}}(\lambda)$]. Below we follow the terminology introduced by Adams [50] and call the waveguide modes guided by both YIG and NC layers as the modes of A-regime, and the modes guided by only one of these layers as the modes of B-regime.

As can be seen from figures 2 (a) - 2(d), the increase of $\Theta$ leads to decrease of both $n_2^{\text{TE}}$ and $n_2^{\text{TM}}$. Varying $\Theta$ one can satisfy one or another condition of Eqs. (14a) – (14c). Thus, for $\Theta = 0.5$ [see figures 2(a)], the condition (14a) is valid for all wavelength range under consideration: 1.0 μm < $\lambda$ < 6.0 μm. The increase of $\Theta$ to 0.62 (figures 2(b)) leads to the intersection of the dispersion curves of $n_2^{\text{TM}}$ and $n_1$ at $\lambda = 1.2$ μm, so that for 1 μm < $\lambda$ < 1.2 μm the condition (14b) takes place, while for 1.2 μm < $\lambda$ < 6.0 μm the condition (14a) is still valid). For $\Theta = 1$ (see figures 2(c)), the condition (14b) is satisfied for 1.2 μm < $\lambda$ < 6.0 μm, while for $\lambda < 1.2$ µm the condition (14c) takes place. The further increase of $\Theta$ to 1.75 ensures the validity of the condition (14c) for all considered wavelength range $\lambda = (1 \div 6)$ µm.

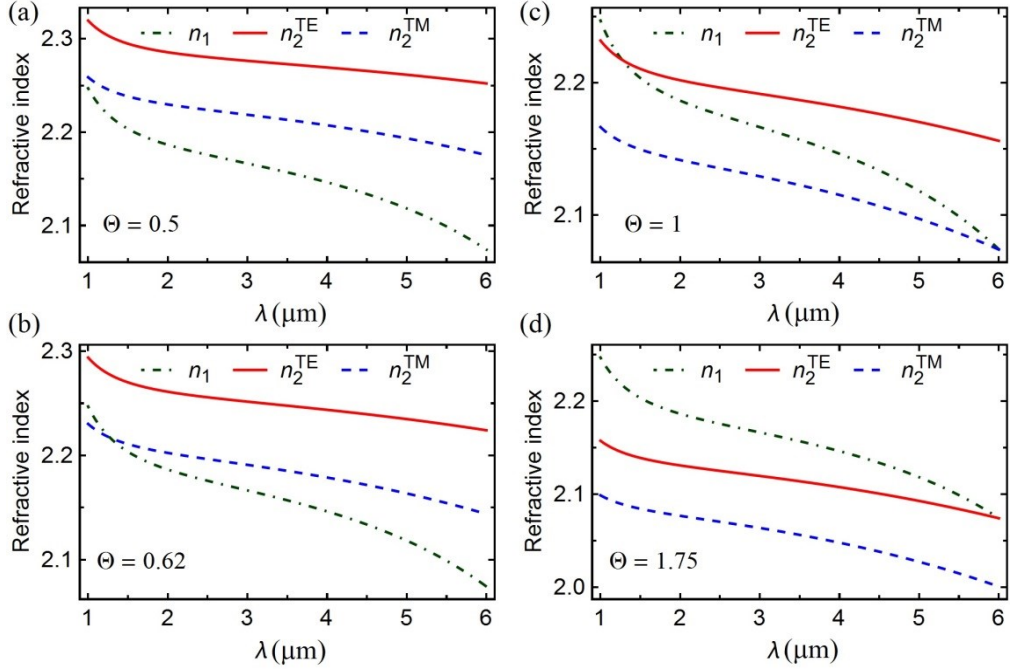

**Figure 2.** Refractive indices dispersion for different values of $\Theta$: 0.5 (a), 0.62 (b), 1 (c), and 1.75 (d). The solid, dashed and dot-dashed lines correspond to $n_2^{TE}(\lambda)$, $n_2^{TM}(\lambda)$ and $n_1(\lambda)$, respectively.

In figure 3 (a) we present the spectra for TE-modes in the wavelength range $\lambda = (1 \div 6)$ µm. The calculations are performed for $L_1 = L_2 = 7$ µm and $\Theta = 1$. The corresponding profile function, *i.e.*, the field distributions within the guiding layers for the fundamental $TE_0$ mode for $\lambda = 1.2$ µm, 1.27 µm and 1.4 µm, are shown in figure 3(b). It should be noted that each dispersion curve is characterized by the corresponding mode number $M$ [see the Eq.(12)], which does not change with the frequency. In figure 3(a), at the fixed frequency $\omega$ one can enumerate the mode from top to bottom: the fundamental mode with $M = 0$ is the highest one (the dashed blue line) and has the largest effective index, the first one has a smaller index, and so on. The dashed lines correspond to the B-regime modes. The modes located in the area $n \in (n_1, n_2)$ (where $n_1 < n_2^{TE}$) propagate in the layer 2 (NC multilayer), while the modes from the region $n \in (n_2, n_1)$ (where $n_2^{TE} < n_1$) propagate in the layer 1 (YIG). The corresponding field profiles are depicted in figure 3(b).

In order to estimate the feasible value of the normalized coefficient $A$ in the Eq.(10), let us take the intensity of the input laser beam to be $I = P/\pi r^2 \approx 10\,\text{W/cm}^2$, where $r = 0.01$ cm is the beam radius and output laser power $P \approx 3$ mW. Considering the geometry of the structure, we assume the input laser beam transforms within the structure into a spot of the elongated ellipse shape with the area $\pi ab \approx \pi r^2$, where $a$ and $b$ are the major (along the $y$-axis) and minor (along the $z$-axis) axes of the ellipse, respectively. Thus, taking $a \approx 0.1$ cm and $b \approx 10$ µm (*i.e.*, guiding film width and thickness), we obtain the value of the power per unit length along the $y$-axis of the order of 1 W/m, which is in agreement with the conventional values (see, e.g. monograph of *A. Yariv* [3]). Thus, for TE-mode the normalization coefficient $A^2 = 8\pi/(c\int_{-\infty}^{\infty} E_y H_z dz)$ is of the order of $10^{-6}$ (erg/cm$^3$).

However, as mentioned before, the asymptotes $n_1$ and $n_2$ can intersect at different values of $\lambda$, depending on $\Theta$. From figure 3(b) and the inset in figure 3(a), one can see that the $TE_0$-mode transforms from the mode, guided by the layer 2 [the dashed blue curve *2* in figure 3(b)], to the mode, guided by the layer 1 [the dashed red curve *1* in figure 3(b)], passing through the area, where it propagates in the regime A, *i.e.*, to the mode, guided by both layers (solid curve *3* in figure 3(b)). The similar situation can be observed for the $TM_0$-mode, but for another value of $\Theta$, as the dispersion curves of $n_2^{TE}$ and $n_2^{TM}$ can intersect the asymptote $n_1$ at the same wavelength (frequency) in the case of the different values of $\Theta$ (see figures 2(b) and 2(c)).

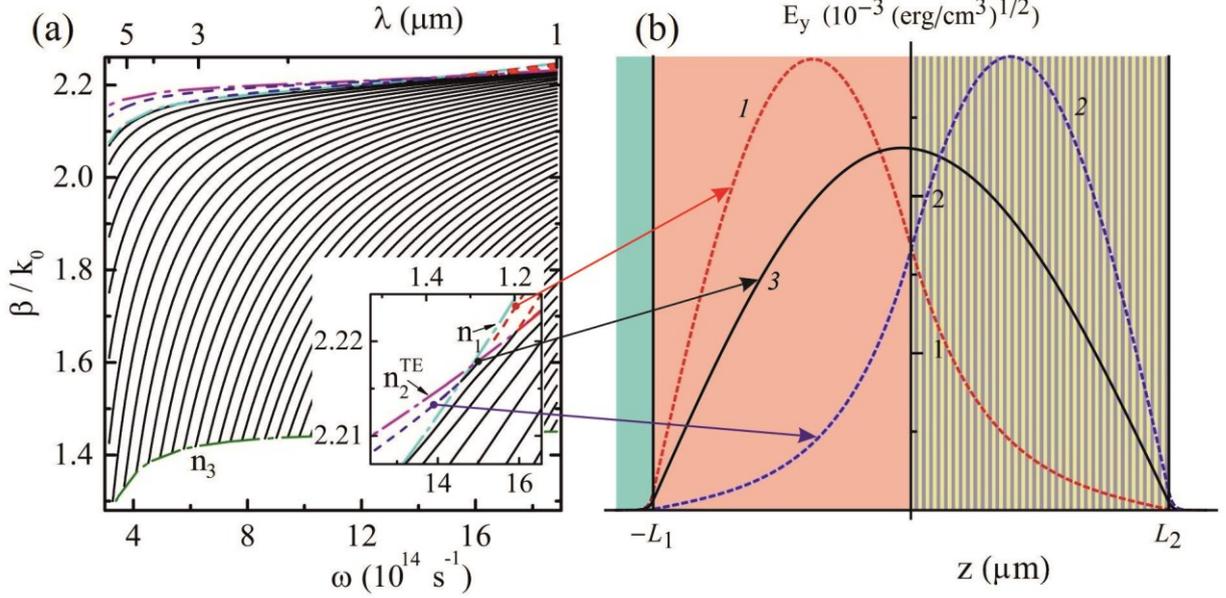

**Figure 3.** (a): the effective mode indices for TE- modes $\beta/k_0$ vs frequency $\omega$; (b): profile functions within the waveguide layers for the fundamental mode TE$_0$ for $\lambda = 1.2$ μm, 1.27 μm and 1.4 μm (the lines marked with numbers 1, 2 and 3, respectively and connected with the corresponding points of the dispersion curves by arrows). The calculations are carried out for $L_1 = L_2 = 7$ μm and $\Theta = 1$.

In figure 4 we demonstrate the dispersion spectra $\beta/k_0$ vs $L_1/\lambda_0$ for TE- (solid red lines) and TM- modes (dashed blue lines) for the cases: (a) $\Theta = 0.3$, (b) $\Theta = 0.8$, (c) $\Theta = 2.0$ calculated for $\lambda_0 = 1.55$ μm, $L_2 = 1.5$ μm. The horizontal dash-dotted lines denote $n_1$, $n_2$ and $n_3$. Depending on the value of $\Theta$ the relation between $n_1$, $n_2^{TE}$ and $n_2^{TM}$ changes [see Eqs. (14a) - (14c)], as depicted in the insets. In the shaded areas in figures 4 and 5, TE- and TM-modes switch from the A- to B-regime, while the modes of both polarizations in the other areas in the figures refer to A-regime only. It means that the shaded areas correspond to the regions where the guiding modes can be localized both within YIG and NC layers (i.e. A-regime) or within only one of the layers (i.e. B-regime). In figure 4(a) one can see, that TE- and TM-modes visibly differ almost in all considered range of $L_1/\lambda_0$. With the increase of the YIG layer thickness the dispersion curves concentration of the A-regime modes takes place (particularly, all modes between the $n_3$ and $n_1$ asymptotes), while the B-regime modes with the effective indices located between the $n_1$ and $n_2^{TE}$ (namely, the TE$_0$-, TE$_1$- and TM$_0$-modes) are not subjected to such a concentration, as one can see in the inset in figure 4(a). Here, in the inset, with the increase of $L_1/\lambda_0$ the dispersion curve for TE$_1$ mode intersects the asymptote $n_1$ and the mode switches from A- to B-regime (i.e. the condition (14a) is satisfied and TE$_1$ mode is now localized in the NC layer).

An important peculiarity of the considered structure is that for $\Theta = 0.8$ and in the vicinity of this value [see figure. 4(b)] the condition $n_2^{TM} < n_1 < n_2^{TE}$ is satisfied, which, in turn, leads to an "inversed" localization of the waves guided in the B-regime: the TE$_0$–mode propagates in the NC-layer, whereas the TM$_0$– and TM$_1$– modes propagate in the YIG-layer. For $\Theta = 2$ [the inset in figure 4(c)] the TE- and TM-modes exhibit a tendency to degeneration (i.e. their dispersion curves almost merge) that is much stronger in the B-regime than in the A-regime. However, unlike the cases depicted in figures 4(a) and 4(b), where only the high-order TE- and TM-modes transfer into the B- regime, here all the modes (both TE- and TM-polarized) convert from the A- to B-regime. Specifically, in the area $\beta/k_0 \in (n_2^{TE}, n_1)$ all modes are the guiding modes of the YIG-layer.

The triangles in figures 4 and 5 (white ones for TE- and black ones for TM-modes) indicate the points where the profile functions (*i.e.*, $E_y(z)$ for TE- and $H_y(z)$ for TM-modes) turn to zero at the interface of the guiding layers ($F_y(0) = 0$). In other words, at these points the electric or magnetic field node pass through the boundary between the guiding layers. The condition of zero field amplitude at the boundary follows from the requirements $C_1, C_2 \to \infty$:

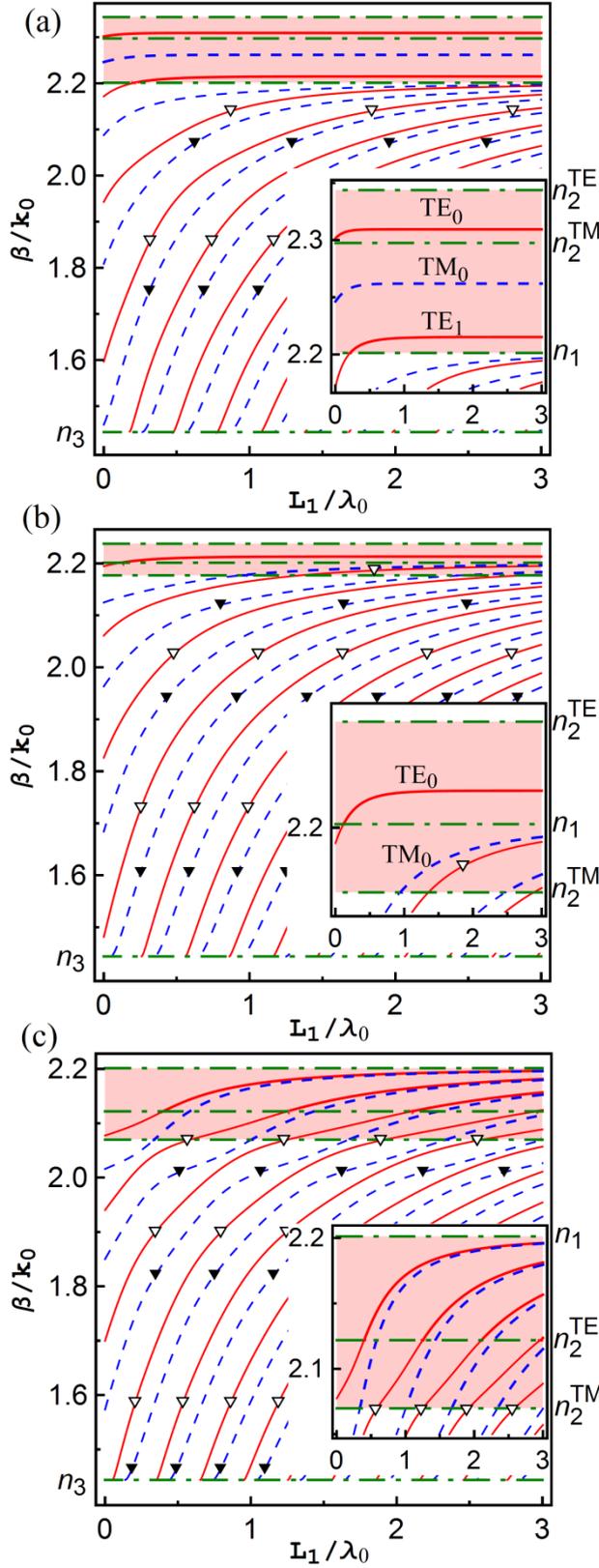
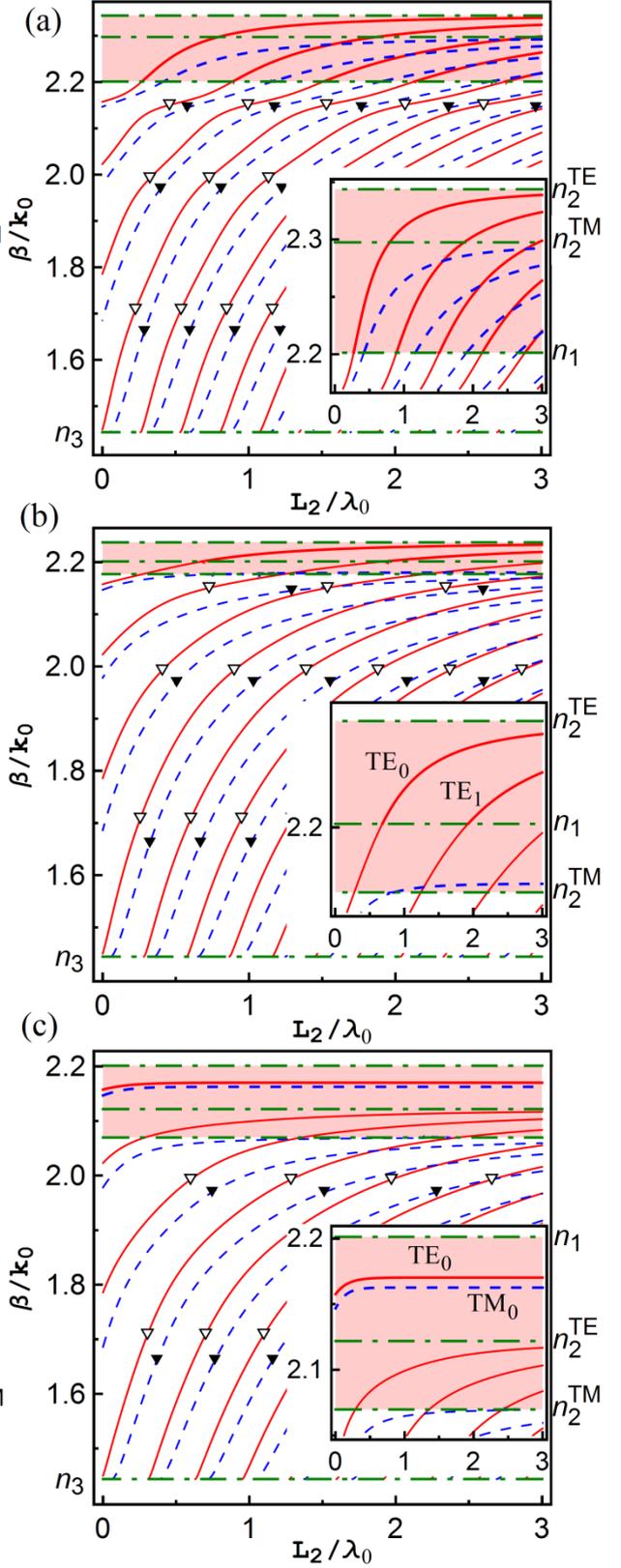

**Figure 4.** Dispersion spectra $\beta/k_0$ vs $L_1/\lambda_0$ for TE- (solid red lines) and TM- modes (dashed blue lines) for the cases: (a) $\Theta = 0.3$, (b) $\Theta = 0.8$, (c) $\Theta = 2.0$ calculated for $\lambda_0 = 1.55\,\mu m$, $L_2 = 1.5\,\mu m$. The dash-dotted lines denote $n_1$, $n_2$ and $n_3$.

**Figure 5.** The same as in figure 4, except for $\beta/k_0$ vs $L_2/\lambda_0$, $L_1 = 1.5\,\mu m$. The non-shaded areas in the figures 4 and 5 refer to the A-regime for the modes of both polarizations.

$$h_1 L_1 + \arctan[h_1 / (\delta p - \beta v)] = \pi l,$$
$$h_2 L_2 + \arctan[h_2 / \sigma q] = \pi l,\qquad(15)$$

where $l = 1, 2, 3...$ is an index of the passing node. The intersections of the solutions of Eqs. (15) with the dispersion curves of the guided modes give the zero field points (the triangles in figures 4 and 5).

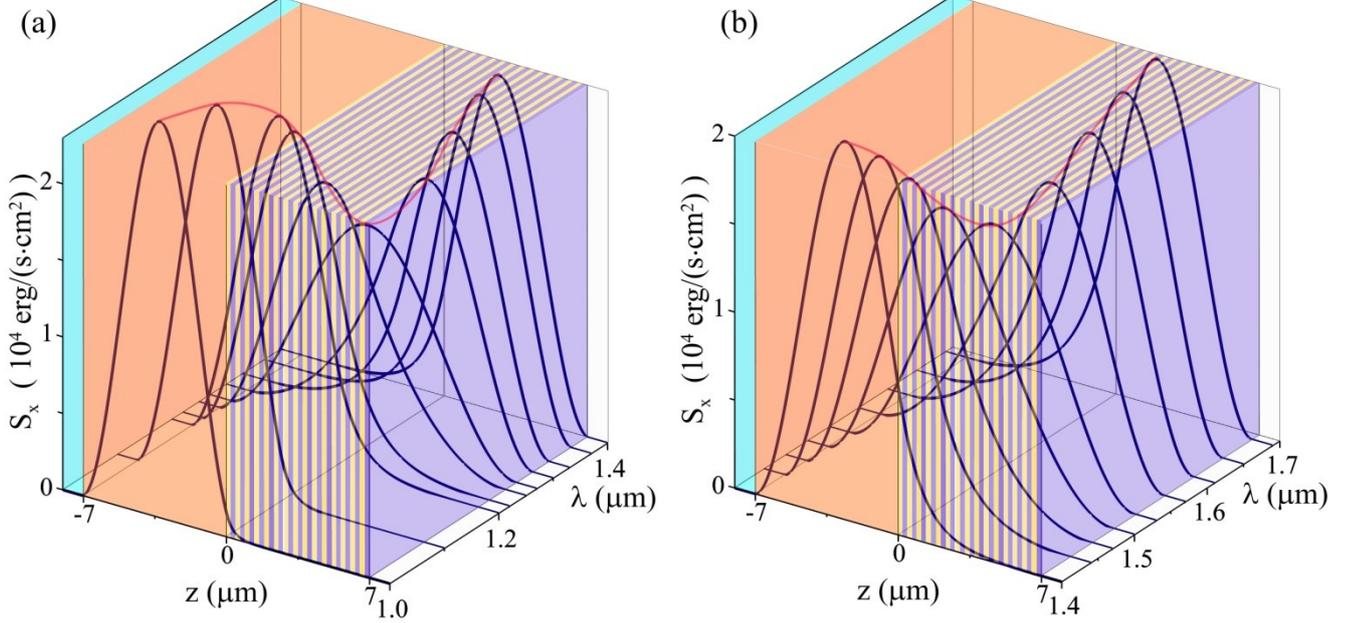

**Figure 6.** Distribution of the longitudinal component of Poynting's vector as a function of wavelength for the fundamental $TE_0$- (a) and $TM_0$-modes (b) for $L_1 = L_2 = 7\,\mu m$, $\Theta = 1.01$ (a) and $\Theta = 0.66$ (b).

It should be mentioned, that the number of nodes in the guiding layer grows with the increase of its thickness [50].

In figure 5 we show the dependence of $\beta / k_0$ on the normalized NC-layer thickness $L_2 / \lambda_0$. The solid red lines and dashed blue lines correspond to TE- and TM-modes, respectively, the dash-dotted lines denote $n_1$, $n_2$ and $n_3$ for the cases: (a) $\Theta = 0.3$, (b) $\Theta = 0.8$, (c) $\Theta = 2.0$. The calculations are carried out for $\lambda_0 = 1.55\,\mu m$ and $L_1 = 1.5\,\mu m$.

Comparing figures 4 and 5, one can observe a significant difference in the modes behavior depending on the type of the layer which thickness is changing. Specifically, with the increase of $L_2 / \lambda_0$ all modes (both TE and TM modes) become the B-regime modes for $\Theta = 0.3$ [see figure 5(a)]. All TE-modes and only one $TM_0$-mode transform to the B-regime for $\Theta = 0.8$ [figure 5(b)], and only the fundamental $TE_0$- and $TM_0$-modes are the modes of the B-regime for $\Theta = 2$ [figure 5(c)]. In the latter case $TE_0$- and $TM_0$–modes remain the B-regime modes independently on $L_2 / \lambda_0$. Moreover, the increase of $L_2 / \lambda_0$ also affects the mode degeneration character: one can observe not the concentration, but the intersections of the dispersion curves for different order modes.

The longitudinal Poynting vector components which characterize the power fluxes across the waveguide structure are given by

$$S_x^{TE} = \frac{c}{8\pi} \mathrm{Re}(E_y H_z),\quad S_x^{TM} = -\frac{c}{8\pi} \mathrm{Re}(H_y E_z) \qquad (16)$$

for TE- and TM- modes, respectively.

Figure 6 illustrates the power flux density redistribution across the structure with the change of the wavelength. For $TE_0$–mode, the flux density peak shifts from the center of YIG layer to the center of NC-multilayer with wavelength increasing from 1.0 μm to 1.4 μm [see figure 6(a)], i.e., the switching between the guiding layers takes place. The envelope curve shows the abrupt lowing of the flux density maximum when the peak reaches the YIG-NC interface. For $TM_0$–mode, the analogous switching takes place at the wavelength interval between 1.4 μm and 1.7 μm and it is more smooth then for $TE_0$–mode [see figure 6(b)].

## 4. Power flux analysis

The spectra of TE-modes (see the inset in figure 3) allow the three different mode propagation regimes within the wavelength regime $\lambda \approx (1.1 \div 1.6)\,\mu m$. So, one can control the electromagnetic wave amplitude in the different guiding layers by adjusting its wavelength, as figure 7 illustrates. The peak of the power flux $P = \int_{-\infty}^{\infty} S_x dz$ moves from the YIG-layer to the NC-multilayer with the increase of the wavelength: the part of the energy carried by the YIG-layer decreases, while that of the NC-layer grows. The partial power fluxes are defined as $P_1 = \int_{-L_1}^{0} S_x dz$ in the YIG-layer and $P_2 = \int_{0}^{L_2} S_x dz$ in the NC-multilayer. The values of the nanolayers thicknesses ratio $\Theta$ are chosen to satisfy the condition of the equal power flux distribution between the waveguiding layers at the wavelength $\lambda = 1.31$ μm for the TE-polarization [figure 7(a)] and at $\lambda = 1.55$ μm for the TM-polarization [figure 7(b)].

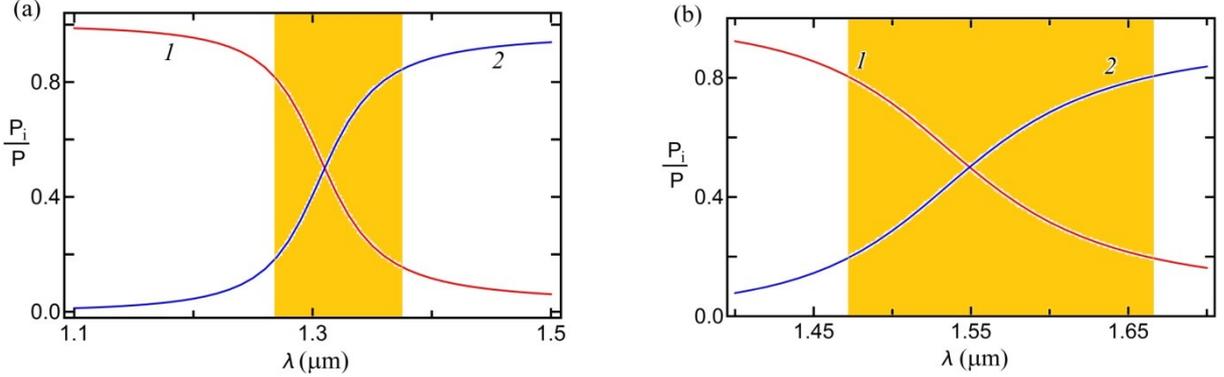

**Figure 7.** The power flux portions $P_i / P$ as a function of wavelength for the fundamental $TE_0$- (a) and $TM_0$-modes (b) (the indices $i = 1, 2$ denote the YIG- and NC-layers); $L_1 = L_2 = 7\,\mu m$, $\Theta = 1.01$ (a) and $\Theta = 0.66$ (b); the shaded areas denote the switching range.

The logarithmic power switching ratio $\eta = 10\log_{10}(P_1 / P_2)$ of the order of 6 dB and higher in the wavelength range of about 100 nm can be achieved for $TE_0$ mode [figure 7(a)]. For $TM_0$ mode [figure 7(b)] a similar switching appears in a twice wider wavelength range (~ 200 nm). Therefore, a wavelength-tunable optical switch can be constructed on the base of the given structure with the possibility of changing the logical state of the waveguide optical cell. Note that for coupling of the waveguide with the laser light into some coupling schemes can be used: end-fire coupling by a lens or by an optical fiber, prism- or grating-assisted coupling (see, e.g., [56]). Moreover, the structure can be used as a two-channel polarization splitter at the fixed $\lambda$, where the $TE_0$-mode propagates in one layer and the $TM_0$-mode propagates in another one.

## 5. Conclusion
In conclusion, we analyzed peculiarities of the waveguide propagation in the four-layer nanocomposite-based MO waveguide structure and demonstrated unique characteristics of its eigenwaves, particularly, polarization switching of TE- and TM-modes depending on the geometrical parameters of the guiding layers. The dispersion relation is obtained taking into account bigyrotropy of the magnetic layer and properties of the nanocomposite layer topology, and an original algorithm of the guided modes identification is proposed.

Our analysis gives better understanding of the evolution of the modes in a four-layer nanocomposite-based bigyrotropic waveguide structures and demonstrates the advantage of the nanocomposite layer in comparison to the homogeneous one in the case of optical switching, opening up a new way forward for the effective using of the similar NC structures. The waveguide propagation regime switching, due to the polarization filtration occurs depending on the geometry parameters (passive guiding light control) of the structure and wavelength (the so-called $\lambda$-tuning).

To summarize, along with low optical losses of the constituting materials at the conventional telecommunications wavelengths the discussed above features of the structure under consideration can find its applications in designing asymmetric and bidirectional optoelectronic and nanophotonic devices such as optical switchers, modulators, isolators which are using for the signal modulation/switching [57,58]. It is also important to note that, considering a gyrotropy of YIG, one can use the structure for an accumulation of the gyrotropy effect in a microresonator photonic cavity.

**Funding.** This research has received funding from the European Union's Horizon 2020 research and innovation programme under the Marie Skłodowska-Curie grant agreement No. 644348 (N.N.D., Yu.S.D., M.K. and I.L.L.), MPNS COST Action MP1403 "Nanoscale Quantum Optics" (N.N.D., Yu.S.D., and I.L.L.), is supported by the grant from Ministry of Education and Science of Russian Federation: Project No. 14.Z50.31.0015 and No. 3.2202.2014/K (N.N.D., Yu.S.D., I.S.P and D.G.S) and also is supported by Ukrainian State Fund for Fundamental Research under project No. Ф71/73-2016 "Multifunctional Photonic Structures" (I.L.L.).